\documentclass[a4paper,11pt]{PoS}

\usepackage{amsmath}
\usepackage[FIGBOTCAP,TABBOTCAP,tight]{subfigure}
\usepackage{tikz}
\usepackage{xcolor}
\usepackage{physics}
\usepackage{lmodern}
\usepackage{dsfont}

\usetikzlibrary{positioning}
\tikzset{
  graph style/.style={
    baseline={([yshift=-.5ex]current bounding box.center)},
    node distance=.5,
  },
  skeleton node/.style={
    black,fill,circle,inner sep=0pt,minimum size=3.5pt
  },
  skeleton bond/.style={
    draw,
  },
}
\makeatletter
\tikzset{
  position/.style args={#1 degrees from #2}{
    at=(#2.#1), anchor=#1+180, shift=(#1:\tikz@node@distance)
  }
}
\makeatother
\title{The Yang-Mills deconfinement transition from a high temperature expansion}

\ShortTitle{The YM deconfinement transition from a HTE}

\author{Jangho Kim, \speaker{Anh Quang Pham}, Owe Philipsen, Jonas Scheunert\\
        Institut f\"ur Theoretische Physik, Goethe-Universit\"at Frankfurt,\\
Max-von-Laue-Str. 1, 60438 Frankfurt am Main, Germany\\
       E-mail:$\hspace{0.09cm}$\email{jkim, pham, philipsen, scheunert@th.physik.uni-frankfurt.de}}

\abstract{The high temperature expansion is an analytical tool to study critical phenomena in statistical mechanics. 
We apply this method to 3d effective theories of Polyakov loops, which have been derived from 4d lattice Yang-Mills by means
of resummed strong coupling expansions. In particular, the Polyakov loop susceptibility is computed as a power series in the effective
couplings. A Pad\'e analysis then provides the location of the phase transition in the effective theory, which can be mapped back
to the parameters of 4d Yang-Mills.  
Our purely analytical results for the critical couplings $\beta_c(N_\tau)$ agree to better than 10\% with those from Monte Carlo simulations.
For the case of $SU(2)$, also the critical exponent $\gamma$ is predicted accurately, 
while a first-order nature as for $SU(3)$ cannot be identified by a Pad\'e analysis. The method can be generalised to include
fermions and finite density.}

\FullConference{37th International Symposium on Lattice Field Theory - Lattice2019\\
		16-22 June 2019\\
		Wuhan, China}

\begin{document}

\section{Introduction}
A lot of effort is dedicated to studying the phase diagram of strongly interacting matter under extreme conditions, which plays an important role for nuclear astrophysics, particle physics and heavy ion collisions. 
Unfortunately, at finite $\mu$ all Monte Carlo algorithms suffer from the "fermion sign problem." Approximate methods work only for small chemical potential $\mu = \mu_B/3 \leq T$ \cite{deForcrand:2010ys}, for which there is no sign of a critical point. Based on stochastic quantization methods, complex Langevin simulations do not have a sign problem, but for complex actions the correct results are not guaranteed \cite{Aarts:2011ax}.

This motivates the development of effective theories. On one hand, with part of the degrees of freedom integrated out, 
the sign problem of effective theories in the cold and dense region of QCD is weaker. 
Thus, it is possible to simulate with standard Monte Carlo or complex Langevin. On the other hand, one can use 
series expansion methods
to carry out analytic computations \cite{Glesaaen:2015vtp}, for which the sign problem is irrelevant. 
In this paper, we present a scheme known as ``high temperature expansion'' in
statistical mechanics, which is also applicable to the effective theories of interest.
As a first test, we apply it to Yang-Mills theory, where we can compare with 
full simulation results.

\section{Effective theories for Yang-Mills}

We start with reviewing the effective theories. Let us consider the $SU(N_c)$ Wilson action which is defined to be
\begin{align}
S(U)=\frac{\beta}{2N_c}\sum_P \left(\Tr U_P +\Tr U_P^\dagger\right)=\sum_P S_P,  
\end{align}where the coupling is $\beta = 2N_c/g^2$, the sum runs over all plaquettes on the lattice and 
\begin{align}
U_P(x)=U_\mu(x)U_\nu(x+\hat{\mu})U_\mu(x+\hat{\nu})^\dagger U_\nu(x)^\dagger.
\end{align}
The associated partition function is given by
\begin{align}
Z = \int \mathcal{D} U \exp[S(U)],
\end{align} where $\mathcal{D} U = \prod_{x, \mu} \mathrm{d}U_\mu(x)$  is the measure and $\mathrm{d}U_\mu(x)$ is the Haar measure on $SU(N_c)$.

We obtain a three-dimensional effective theory via integrating out all spatial links $U_i(x)$, which was first done for $SU(2)$ YM theory \cite{Polonyi:1982wz},
\begin{align}
Z &= \int \mathcal{D}U_0 \exp[S_{\mathrm{eff}}(U_0)],\\
S_{\mathrm{eff}}(U_0)&= \ln\int\mathcal{D}U_i \exp[S(U)] = \lambda_1 S_1 
	+ \lambda_2 S_2+\ldots\;.
\end{align} 
Here $\lambda_n(N_{\tau}, \beta)$ denote the effective couplings, which are functions of the temporal extent $N_\tau$ and the original coupling $\beta$. 
The $S_i$ depend only on the temporal links $U_0(x)$ and for $N_c\leq3$ the
dependency can be re-expressed in terms of Polyakov loops
\begin{align}
L_{\vec{x}} = \Tr W_{\vec{x}} = \Tr \prod_{\tau=0}^{N_{\tau}-1} U_0(\tau,\vec{x}).
\end{align}

In practice, the effective action is obtained by a character expansion of the
gauge-invariant function $\exp(S_P)$, and for the orders considered here the
integrations over the spatial links can be done using the orthogonality
relations of characters. (For details on the
character expansion, see \cite{Drouffe:1983fv}). This results in a power series
in the coefficient of the fundamental character, $u(\beta)$, which is always
smaller than one for finite $\beta$.

Here we work with the nearest neighbour effective action derived in
\cite{Langelage:2010yr} which includes a logarithmic resummation and reads
(since we only deal with one coupling we drop the index on $\lambda_1$ from now
on)
\begin{equation}
  \label{eq:efft-nn}
  S_{\mathrm{eff}}
  =
  \begin{cases}
    \sum_{\langle\vec{x},\vec{y}\rangle}
    \ln[1+\lambda(N_\tau,u) L_{\vec{x}}L_{\vec{y}}],
    & \text{ for } N_c = 2, \\
    \sum_{\langle\vec{x},\vec{y}\rangle}
    \ln[1+\lambda(N_\tau,u) (L_{\vec{x}}L_{\vec{y}}^* +L^*_{\vec{x}}L_{\vec{y}})],
    & \text{ for } N_c = 3,
  \end{cases}
\end{equation}
with the effective coupling
\begin{align}
  \lambda(u,N_\tau\geq 5) 
  & =
  u^{N_\tau}
  \exp\left[
    N_\tau\left(4 u^4-4u^6+\frac{140}{3}u^8-\frac{36044}{405}u^{10}\right)
  \right],
  & \text{ for } N_c = 2, \\
  \begin{split}
  \label{eq:map-ld}
    \lambda(u,N_\tau\geq 6)
    & = 
    u^{N_\tau}
    \exp\biggl[
      N_\tau(
      4u^4 + 12u^5 - 14u^6 - 36u^7 + \\
      & \hspace{80pt}
      \frac{295}{2}u^8 + \frac{1851}{10}u^9 + \frac{1055797}{5120}u^{10})
    \biggr],
  \end{split}
  & \text{ for } N_c = 3.
\end{align}

\section{High temperature expansion for the effective YM theory}

To solve the effective theories, we used high temperature
expansions\footnote{High temperature expansion is the name adapted from
statistical mechanics, in context of effective theories we expand our theories
around small couplings $\lambda$.} which played an essential role for studying
critical phenomena before the Monte-Carlo method. 
General concepts of the high temperature expansion applied to Ising models can
be found, for instance, in \cite{Domb:1974wz}. This method is particularly
simple for Ising models because spins can only take two values $+1$ or $-1$,
but for effective gauge theories it is more complicated when our ``spins''
are Polyakov loops. 

Let us focus on the improved effective action \eqref{eq:efft-nn} for $N_c=3$,
the $N_c=2$ can be treated analogously with the simplification that the
Polyakov loop is always real.
Our main goal is to extract information about phase transitions from the
series. Thus, we are interested in the Polyakov loop susceptibility $\chi_L$
which peaks near phase transitions
\begin{align}
\label{eq:chi}
\chi_L = \frac{1}{V} \frac{\partial^2}{\partial J^2}\ln Z[J]\Bigg|_{J=0},
\end{align}
where $Z[J]$ is the partition function coupled to an external source $J$,
\begin{align}
Z[J] = 
\int \mathcal{D}\mathbb{U}[J] \prod_{\langle\vec{x},\vec{y}\rangle}[1+\lambda (L_{\vec{x}}L_{\vec{y}}^* +L^*_{\vec{x}}L_{\vec{y}})]=
	\int \prod_{\vec{x}} \mathrm{d}W_{\vec{x}}e^{J(L_{\vec{x}}+L^*_{\vec{x}})} \prod_{\langle\vec{x},\vec{y}\rangle}[1+\lambda 
	(L_{\vec{x}}L_{\vec{y}}^* +L^*_{\vec{x}}L_{\vec{y}})]\;.\nonumber
\end{align}
We expand the partition function in the effective coupling $\lambda$,
\begin{align}
Z[J]
    & =
      \int \mathcal{D}\mathbb{U}[J]
       +
      \lambda \int  \mathcal{D}\mathbb{U}[J]
      \sum_{\left<\vec{x},\vec{y}\right>}
      2 \Re(L_{\vec{x}}L_{\vec{y}}^\ast)\nonumber
       \\
    & \quad +
      \lambda^2 \int  \mathcal{D}\mathbb{U}[J]
      \sum_{\left<\vec{w},\vec{x}\right>}
      \sum_{\left<\vec{y},\vec{z}\right>}
      4 \Re(L_{\vec{w}}L_{\vec{x}}^\ast) \Re(L_{\vec{y}}L_{\vec{z}}^\ast) +
      \mathcal{O}(\lambda^3).
	\label{eq:exp}	
\end{align}
Fortunately, instead of lengthy and complicated expressions there is an elegant 
graph representation for (\ref{eq:exp}),
\begin{align}
Z[J]
    & = \;
      \begin{tikzpicture}[graph style]
        \node[skeleton node] (n1) {};
      \end{tikzpicture} \; + \lambda \;
      \begin{tikzpicture}[graph style]
        \node[skeleton node] (n1) {};
        \node[skeleton node] (n2) [right=of n1] {};
        \draw[skeleton bond] (n1) -- (n2);
      \end{tikzpicture} + \lambda^2 \; \Bigl(\;
      \begin{tikzpicture}[graph style]
        \node[skeleton node] (n1) {};
        \node[skeleton node] (n2) [position=60 degrees from n1]  {};
        \node[skeleton node] (n3) [position=300 degrees from n2] {};
        \draw[skeleton bond] (n1) -- (n2);
        \draw[skeleton bond] (n2) -- (n3);
      \end{tikzpicture} \; + \;
      \begin{tikzpicture}[graph style]
        \node[skeleton node] (n1) {};
        \node[skeleton node] (n2) [right=of n1] {};
        \node[skeleton node] (n3) [below=of n1] {};
        \node[skeleton node] (n4) [right=of n3] {};
        \draw[skeleton bond] (n1) -- (n2);
        \draw[skeleton bond] (n3) -- (n4);
      \end{tikzpicture} \;
      \Bigr) + \mathcal{O}(\lambda^3).
	\label{eq:graph}	
\end{align} 
Here the order of $\lambda$ corresponds to the number of bonds of an associated
graph, and from the number of bonds coupled to a vertex one can count the
number of Polyakov loops at this vertex. 
Note that for our effective theories we need to include open graphs, for
instance at $\lambda^2$, while those contributions drop out in most cases for
Ising models. There are two main computations to be performed:
\begin{itemize}
\item $\textbf{Graph counting.}$ In general, each graph has its own embedding number, 
	which gives the number of ways this graph appears on the lattice.  
		For a detailed discussion we refer to \cite{Domb:1974wz}. 
		On a three dimensional cubic lattice the embedding numbers of each 
		graph in (\ref{eq:graph}) are
\begin{align*}
V,\hspace{0.3cm} 3V,\hspace{0.3cm} 15V, \hspace{0.3cm} 9V^2/2 - 33V/2,
\end{align*} where $V=N_s^3$ is the number of lattice sites. 
\item $\textbf{Graph evaluation.}$ As an example, the fist order graph of $Z[J]$ is calculated explicitly as follows
  \begin{align*}
    \begin{tikzpicture}[graph style]
      \node[skeleton node] (n1) {};
      \node[skeleton node] (n2) [right=of n1] {};
      \draw[skeleton bond] (n1) -- (n2);
    \end{tikzpicture}
    &=
    \int \mathcal{D}\mathbb{U}[J]
    \sum_{\left<\vec{x},\vec{y}\right>} 2 \Re(L_{\vec{x}}L^*_{\vec{y}})\\
    &=
    3V\left(\int \mathrm{d}\mathbb{U}[J]\ \right)^{V-2} 
    2\left(\int \mathrm{d}\mathbb{U}[J]\ L \right)
    \left(\int \mathrm{d}\mathbb{U}[J]\ L^*\right),
  \end{align*}
The nearest neighbor sum gives $3V$ integrals of the same kind, i.e., it
reduces our computations to single site $SU(N_c)$ integrals as one can see in
the second equation. Finally, it is easy to compute $\chi_L$ with use of \eqref{eq:chi}
and the group integrals at the sites 
\begin{align}
I(n,m) = \int_{SU(N_c)} \mathrm{d} W L^n L^{*m},
\end{align}
can be evaluated for $N_c=3$ using the formula given in \cite{GATTRINGER2011242}.
This formula is based on a generating functional derived in \cite{Uhlmann:2006ze}.
This reference also contains a generating functional for the $SU(2)$ case,
which one can handle a similar fashion.
\end{itemize}
It is obvious that beyond the first few orders the computations are cumbersome,
but the graphs can be generated, embedded and evaluated by a computer. The
original algorithm to compute the embedding constants was developed by Martin
\cite{Martin:1974wz}, here we use the code provided in \cite{Hamer:2006zz}.

\section{Series analysis and results}

Phase transitions are signalled by non-analyticities in the susceptibility. In particular,
near a critical point the susceptibility diverges with a critical exponent,
such that its log-derivative has a simple pole,
\begin{align}
\chi_L\sim\frac{1}{(\lambda - \lambda_c)^\gamma}\hspace{0.5cm}\longrightarrow \hspace{0.5cm} \frac{\mathrm{d}}{\mathrm{d}\lambda}\ln \chi_L \sim\frac{-\gamma}{\lambda-\lambda_c}.
\end{align}
At a first-order phase transition, instead,
there is a discontinuity at the critical coupling.
The generic expression obtained by expansion methods is a finite 
power series in $\lambda$,
\begin{align}
	\chi_L = \sum_{n=0}^{N+1} s_n\lambda^n,
\end{align}
which cannot show such behaviour.
\begin{figure}[t]%
    \centering
    \subfigure[$SU(3)$]{\label{fig:su3crilam} \includegraphics[width=0.46\textwidth]{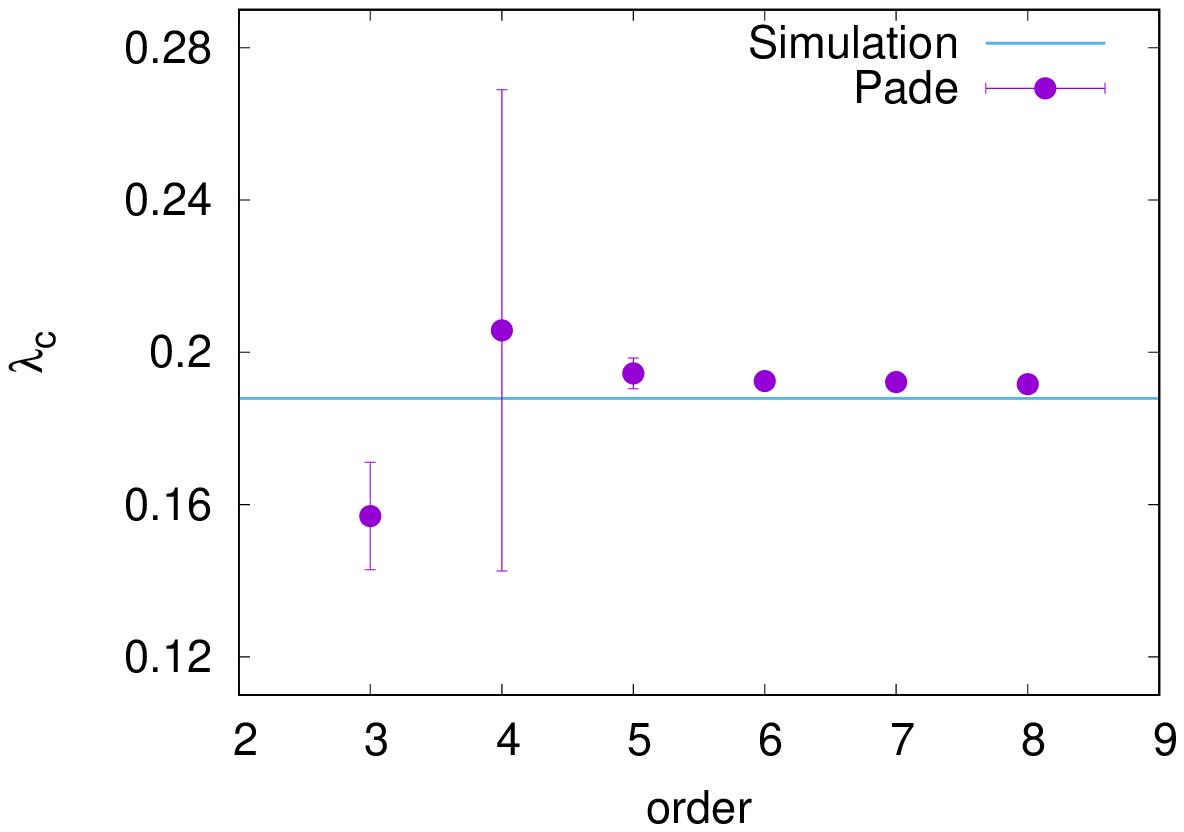}}%
    \qquad
    \subfigure[$SU(2)$]{\label{fig:su2crilam} \includegraphics[width=0.46\textwidth]{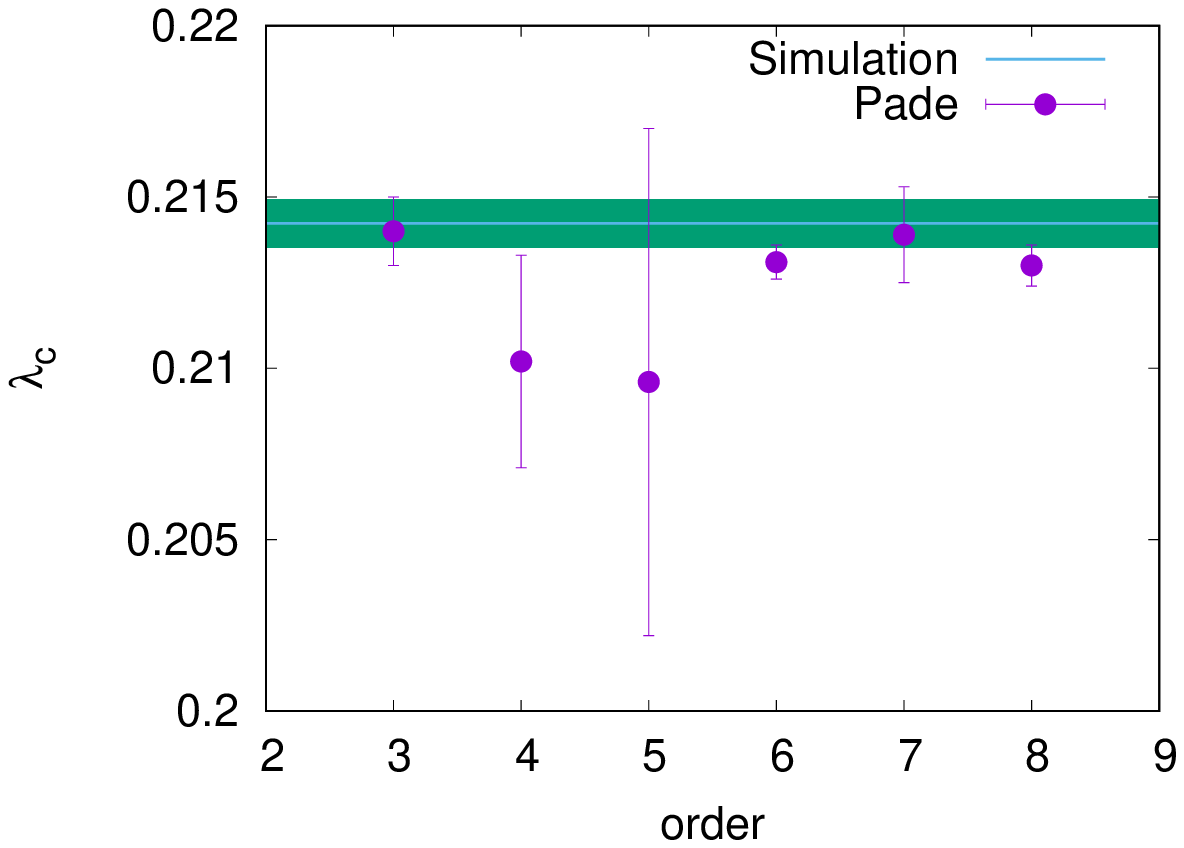}}
    \caption{\label{fig:example} Critical coupling $\lambda_c$ to different orders}%
\end{figure}

A standard method to extract information about criticality from the available
coefficients is the Pad$\acute{\text{e}}$ analysis \cite{Guttmann1989wz}. 
Finite series can be extended 
to infinite order by constructing ratios of two polynomials of degrees $L,M$, 
whose expansion agrees with the original series to its given order $N$.
We then approximate
\begin{align}
\frac{\mathrm{d}}{\mathrm{d}\lambda}\ln \chi_L \sim[L/M]=P(\lambda)/Q(\lambda)=\frac{p_0+p_1\lambda+\dots+p_L\lambda^L}{1+q_1\lambda+\dots+p_M\lambda^M},
\end{align}in which the $L+M+1=N+1$ coefficients $p_0,p_1,\dots,p_L,q_1,\dots,q_M$ 
are determined by equating like powers of $\lambda$.
Diagonal Pad$\acute{\text{e}}$ approximants are expected to be the most stable ones,
since they are
invariant under M\"obius transformations and thus under resummations of the series.
We extract our results from
the table of diagonal and near-diagonal approximants $[M/M]$, $[M-1,M+1]$ and $[M+1/M-1]$. 
At order $\lambda^N$ in \eqref{eq:log} we take either
three or two approximants depending on whether $N$ is even or odd, e.g.~,
$([1/2], [2/1])$ at order 3 and $([1/3],[2/2],[3/1])$ at order 4. For each of these
we determine the poles in $\lambda$ and their residues.
We then compute the mean value and quote as systematic error
the maximum deviation between pairings $|\lambda_j - \lambda_i|$, where $i$
labels the approximants.

$\textbf{Results.}$ Our result for the logarithmic derivative of the Polyakov loop susceptibility up to 8th order reads 
\begin{align}
\frac{\mathrm{d}}{\mathrm{d}\lambda}\ln \chi_L = 6&+24\lambda+126\lambda^2+768\lambda^3+3126\lambda^4+20736\lambda^5+87198\lambda^6\nonumber\\
&+\frac{9489884859334657}{17179869184}\lambda^7+\frac{83864630933323767}{34359738368}\lambda^8+\dots
	\label{eq:log}
\end{align}
Note that the coefficients of (\ref{eq:log}) at order seven and higher become 
rational due to disconnected contributions. 
It is now straightforward to estimate the critical coupling $\lambda_c$ and the 
critical exponent $\gamma$ by the Pad$\acute{\text{e}}$ analysis. Note however,
that Pad\'e approximants always produce poles, even in the case of a first-order 
transition. In this case they signal the end of the metastability region
and thus are an upper bound on the true critical coupling. For an example,
see the $q$ state Potts model \cite{PhysRevE.74.051113}. 

In Figure~\ref{fig:su3crilam} we show the critical coupling $\lambda_c$ for $N_c=3$ to each order 
of \eqref{eq:log}. 
The best estimate of $\lambda$ is within $2\%$ of the simulation result for that model \cite{Langelage:2010yr}. 
Note that, even though the results have stabilised with inreasing order, they slightly overshoot the true 
value. This is due to the first-order nature of the transition, where the series probes the end of the metastability region.
The same computations for the case $N_c=2$ are shown in Figure~\ref{fig:su2crilam}. Here the 
best estimate for $N_c=2$ is within $1\%$ from simulation data. 
 
\begin{figure}[t]%
    \centering
    \vspace*{-0.5cm}
    \subfigure[]{
      \label{fig:EtvsMC} 
      \includegraphics[width=0.47\textwidth]{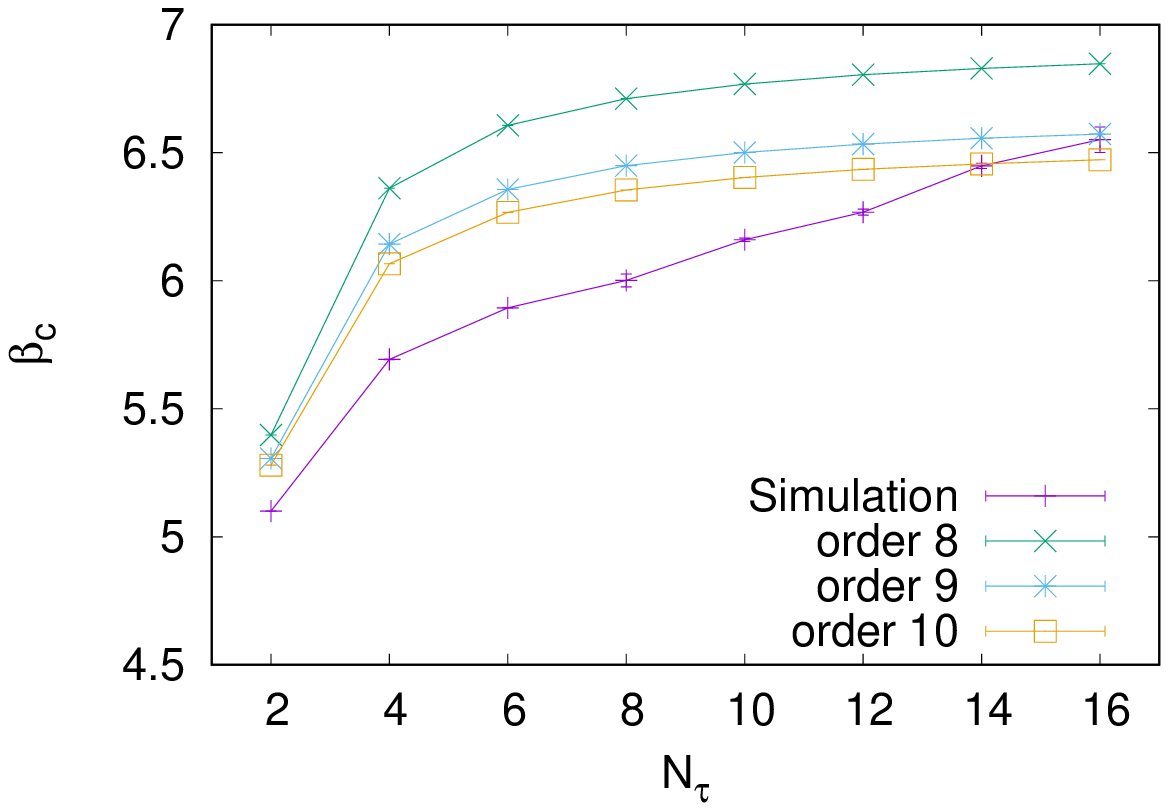}
    }
    \subfigure[]{
      \label{fig:cri_su2_su3} 
      \includegraphics[width=0.47\textwidth]{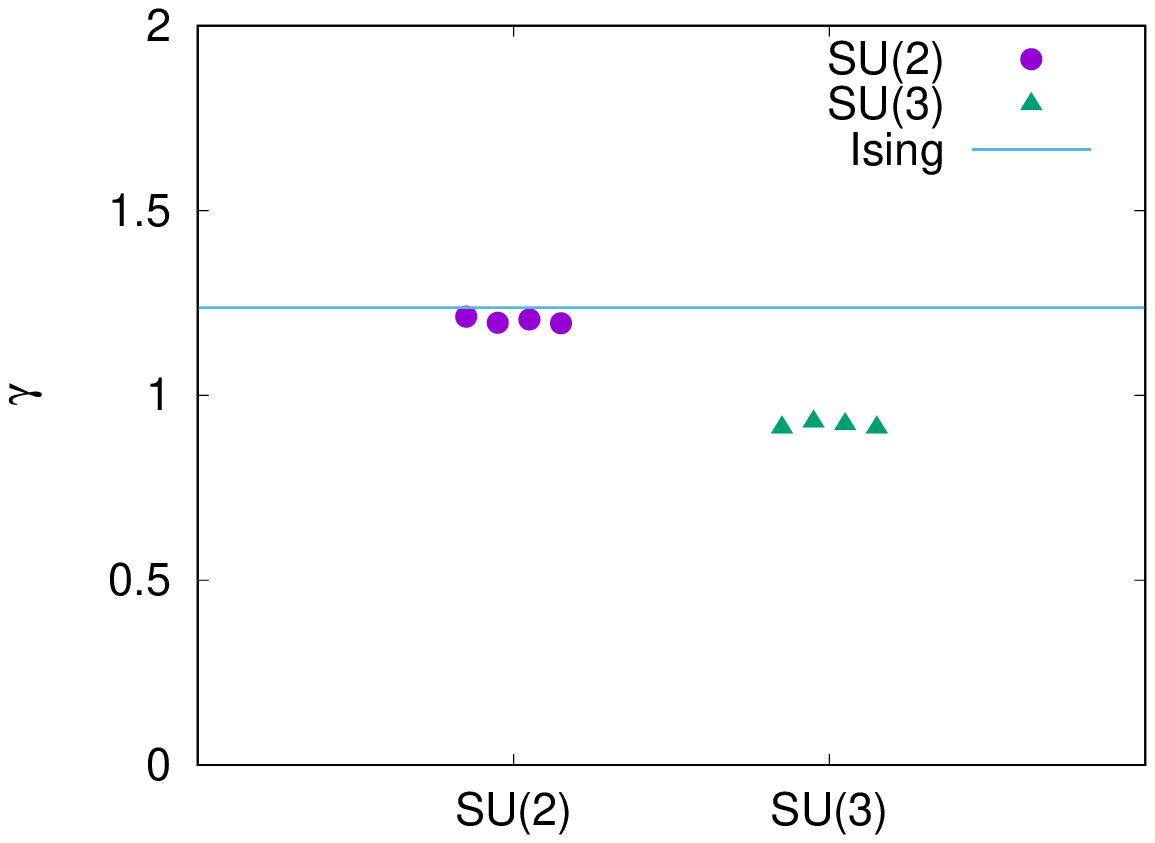}
    }
    \caption{\label{fig:example-2} (a) Critical coupling $\beta_c$ by \eqref{eq:map-ld} for $SU(3)$. 
    (b) Exponent $\gamma$ for $SU(2)$ and $SU(3)$. 
	 }%
\end{figure}

We can now invert \eqref{eq:map-ld} to extract the critical couplings $\beta_c$  
and compare with results of the full four-dimensional Yang-Mills theory
\cite{Fingberg:1992ju}. Figure~\ref{fig:EtvsMC} shows this for $SU(3)$ and we
observe a satisfactory convergence behavior of strong coupling series to the
full result.

In Figure~\ref{fig:cri_su2_su3}, by using four different Pad$\acute{\text{e}}$ approximants $[4/4]$, $[4/3]$, $[3/3]$, and $[3/5]$ we 
show the resulting critical exponents
$\gamma$ of our effective theory. 
The estimates for $SU(2)$ accurately reproduce the value for the 
universality class of the $3d$ Ising model, 
while the $SU(3)$ case gives values different from all known 
universality classes. This is consistent with its nature as a first order phase transition, see the discussion above.

\section{Conclusions}
We have considered a 3d effective theory for finite temperature $SU(2),\, SU(3)$ Yang-Mills in 4d, obtained by strong coupling methods,
which only depends on  Polyakov loops. 
We computed the Polyakov loop susceptibility as a series in the effective coupling and, by means of a Pad\'e analysis, 
extracted critical couplings $\beta_c$ for the phase transition by purely analytical methods. The results agree  
to better than 10\% with simulations for $N_\tau=2-16$. For $SU(2)$, also the critical exponent $\gamma$ 
is accurately reproduced, while Pad\'e analyses generally cannot identify a first-order transition.
This work demonstrates the benefit of going via an effective theory, compared to a strong coupling expansion of the 
full theory \cite{Langelage:2009jb}, and can be straightforwardly extended to include fermions at finite density.

\acknowledgments
This work is supported by the Deutsche Forschungsgemeinschaft (DFG) through the grant CRC-TR 211 "Strong-interaction matter under extreme conditions" and by the Helmholtz International Center for FAIR within the LOEWE program of the State of Hesse.

\bibliography{ref}

\end{document}